\documentclass[reprint,aps,prl,twocolumn,superscriptaddress]{revtex4-1}
\usepackage{braket}
\usepackage{amssymb}
\usepackage{amsmath}
\usepackage{epsfig}
\usepackage{color}
\usepackage{graphics, graphicx}
\usepackage{bbold}
\usepackage{psfrag}
\usepackage{mathcomp}
\usepackage{subfigure}
\usepackage{verbatim}
\usepackage[colorlinks, citecolor=blue]{hyperref}
\usepackage[normalem]{ulem}
\usepackage{bm}
\def\cp#1{\mathbf{#1}}
\begin{document}

\title{Shell-Shaped Quantum Droplet in a Three-Component Ultracold Bose Gas}
\author{Yinfeng Ma}
\affiliation{Beijing National Laboratory for Condensed Matter Physics, Institute of Physics, Chinese Academy of Sciences, Beijing 100190, China}
\affiliation{Department of Basic Courses, Naval University of Engineering, Wuhan 430033, China}
\author{Xiaoling Cui}
\email{xlcui@iphy.ac.cn}
\affiliation{Beijing National Laboratory for Condensed Matter Physics, Institute of Physics, Chinese Academy of Sciences, Beijing 100190, China}

\date{\today}

\begin{abstract}
Shell-shaped Bose-Einstein condensate (BEC) is a typical quantum system in curved geometry. Here we propose a new type of shell-shaped BEC with self-bound character, thereby liberating it from stringent conditions such as microgravity or fine-tuned trap.  Specifically, we consider a three-component (1,2,3) ultracold Bose gas where (1,2) and (2,3) both form quantum droplets. The two droplets are mutually immiscible due to strong 1-3 repulsion, while still linked by component-2 to form a globally self-bound object. The outer droplet then naturally develops a shell structure without any trapping potential.  It is shown that the shell structure can significantly modify the equilibrium density of the core, and lead to unique collective excitations highlighting the core-shell correlation.   All results have been demonstrated in a realistic $^{23}$Na-$^{39}$K-$^{41}$K mixture.  
By extending quantum droplets from flat to curved geometries, this work paves the way for future exploring the interplay of quantum fluctuations and non-trivial real-space topologies in ultracold gases.
\end{abstract}
\maketitle

Quantum systems in curved geometries exhibit many distinctive features due to their non-trivial real-space topologies. For instance, a periodic boundary allows a persistent superflow of toroidal Bose-Einstein condensates (BECs)~\cite{ring_1, ring_2}, and a local curvature gives rise to new topological defects\cite{Nelson, Ho, Zhou}, interesting few-body physics\cite{Zhang_Ho, Shi_Zhai} and even non-Hermitian phenomena\cite{ZhouQi}. 
As an outstanding case with non-trivial real-space topology, the shell-shaped BEC has attracted a great amount of attention in the field of ultracold atoms\cite{review}, and various fascinating properties have been revealed in terms of ground state and thermodynamics\cite{Tononi,Rhyno,ciardi,prestipino},  collective modes\cite{Lannert,Padavic,Sun}, expansion dynamics\cite{Tononi2} and vortex formation\cite{Padavic2,Fetter}. These studies are closely related to the experimental efforts in creating shell-shaped BECs, or atomic bubbles, using the shell-shaped potentials under  radio-frequency dressing\cite{proposal_rf_dress_1,proposal_rf_dress_2}. However, the Earth's gravity  prevents the formation of a closed shell in such setup\cite{Colombe}, and consequently a microgravity environment is required. Indeed, the first shell-shaped BEC was recently  realized in NASA Cold Atom Laboratory aboard the 
International Space Station\cite{NASA_expt}. Shortly after that, it was proposed alternatively that binary bosons in immiscible regime\cite{Ho_Shenoy, Pu_Bigelow} can also achieve the shell structure\cite{Wolf}. 
This idea has been  successfully implemented in a recent experiment in the presence of Earth's gravity\cite{Dajun_expt}, where a magic-wavelength optical trap  was applied to avoid different gravitational sags between two species. 

\begin{figure}[t]
    \centering
    \includegraphics[width=8.5cm]{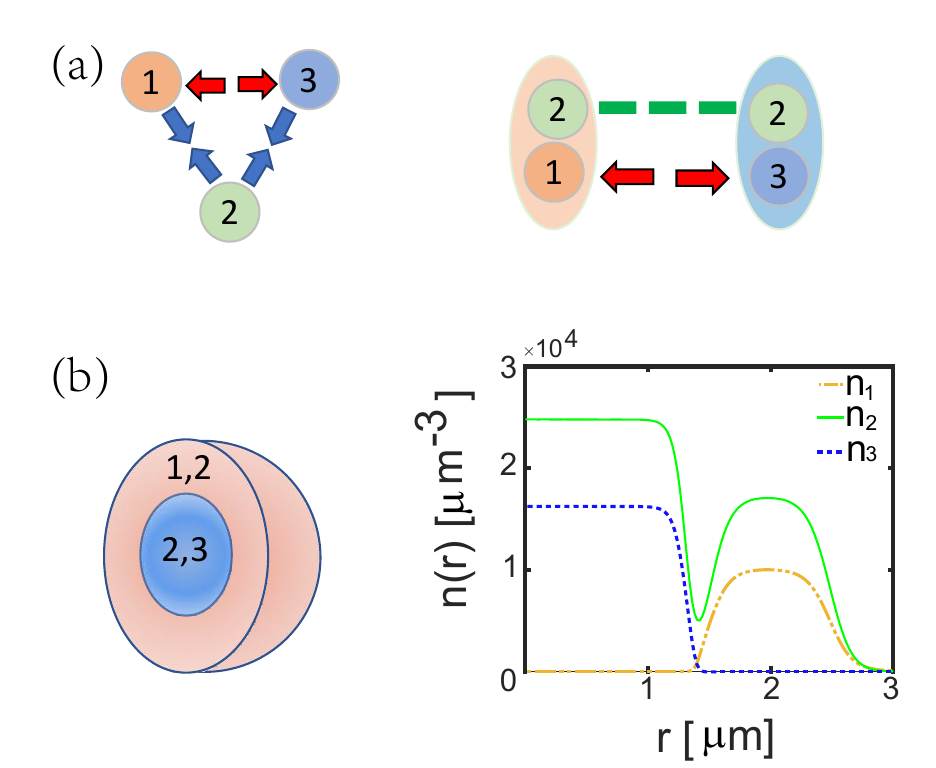}
    \caption{(a) Interaction model for realizing shell-shaped geometry in three-component('1', '2', '3') bosons. Here (1,2) and (2,3) both form quantum droplet due to inter-species attraction, while the two droplets are immiscible due to  1-3 repulsion. The outer droplet is then shell-shaped. The common component-2 links the whole system together as a self-bound object. (b) Typical half-sphere distribution and according density profile along the radius direction. Here we consider a realistic $^{23}$Na-$^{39}$K-$^{41}$K ('1'-'2'-'3') mixture near $B\sim 150$G with  $a_{23}=-70a_0$ and $(N_1,N_2,N_3)/10^5=(5,10.7,1.5)$. }\label{fig_schematic}
\end{figure}

In this work, we introduce a new type of shell-shaped BEC without resorting to any trapping potential, and therefore it does not rely on microgravity environment or fine-tuned traps as in previous experiments\cite{Dajun_expt, NASA_expt}. Our scheme is motivated by the recent development of quantum droplet in ultracold atoms, which has been realized in both dipolar gases\cite{ferrier-barbut_observation_2016,schmitt_self-bound_2016,ferrier-barbut_liquid_2016,chomaz_quantum-fluctuation-driven_2016,tanzi_observation_2019,bottcher_transient_2019,Chomaz_long_lived_2019} and boson mixtures\cite{cabrera_quantum_2018,cheiney_bright_2018,semeghini_self-bound_2018,derrico_observation_2019,burchianti_dual-species_2020,guo_lee-huang-yang_2021}. These droplets are stabilized by a mean-field attraction and the Lee-Huang-Yang repulsion from quantum fluctuations\cite{Petrov}, and thus can be self-bound in vacuum. 
Here we remark that the self-bound nature of quantum droplet offers an ideal opportunity for creating perfect shell geometry on Earth. This is because in the absence of any external trap, different species in the droplet will fall freely with the same speed and therefore no relative displacement will be produced due to the gravity.

Our scheme of creating shell-shaped BEC is illustrated in Fig.\ref{fig_schematic}. Specifically, we consider a three-component (1,2,3) boson mixture with contact interactions. Here (1,2) and (2,3) both form quantum droplets due to inter-species attractions, while they are immiscible due to strong 1-3 repulsion. The shared component-2 acts as a glue to link two droplets together as a globally self-bound object. In this way, the outer droplet is repelled by the inner core and naturally develops a shell structure in free space. Using a realistic mixture of $^{23}$Na-$^{39}$K-$^{41}$K near  $B\sim150$G, we show that  the outer shell can efficiently expanded to larger radius with thinner width by increasing the core size. To balance with the outer shell, the core droplet exhibits very different equilibrium densities from the vacuum case (without shell). The core-shell correlation can also be manifested in the unique collective excitations of the system. 
As the first study of quantum droplet in curved geometry, our work opens a new avenue for exploring the interplay of quantum fluctuations and non-trivial real-space topologies in the platform of ultracold atoms. 

We write down the Hamiltonian of three-component boson mixtures  $H=\int d{\bf r} H({\bf r})$, with ($\hbar=1$)
\begin{equation}
H({\bf r})=\sum_{i=1}^3\phi_{i}^\dagger({\bf r})\left(-\frac{\nabla^2}{2m_i}\right)\phi_i({\bf r})+\sum_{ij}\frac{g_{ij}}{2}\phi_{i}^\dagger\phi_{j}^\dagger\phi_{j}\phi_{i}({\bf r}).
\end{equation}
Here ${\bf r}$ is the coordinate; $m_i$ and $\phi_{i}$ are the mass and field operator of boson species $i$, respectively; $g_{ij}=2\pi a_{ij}/m_{ij}$ is the coupling constant between species $i$ and $j$, with scattering length $a_{ij}$ and reduced mass  $m_{ij}=m_im_j/(m_i+m_j)$. 

For a homogeneous dilute gas with densities $\{ n_i\}$ $(i=1,2,3)$, the total energy density is composed by the mean-field part  $\epsilon_{\rm mf}=\frac{1}{2}\sum_{ij}g_{ij}n_in_j$ and a correction from quantum fluctuations: 
\begin{equation}
\epsilon_{\rm qf}=\int \frac{d^3{\bf k}}{2(2\pi)^3}\left[\sum_i(E_{i{\bf k}}-\epsilon_{i{\bf k}}-g_{ii}n_i)+\sum_{ij}\frac{2m_{ij}g_{ij}^2n_in_j}{{\bf k}^2}\right] \label{E_qf}
\end{equation}  
with $\epsilon_{i{\bf k}}=k^2/2m_{i}$, and $E_{i{\bf k}}$ the $i$-th Bogoliubov mode\cite{ma_borromean_2021}. 
For a general system with inhomogeneous densities, we employ an extended Gross-Pitaevskii(GP) equations incorporating quantum fluctuations: 
\begin{equation}
\begin{split}
i\partial_t\phi_i=\left(-\frac{\nabla^2}{2m_i}+\sum_{j}g_{ij}n_j+\frac{\partial \epsilon_{\rm qf}}{\partial n_i}\right)\phi_i. 
\end{split} \label{GP}
\end{equation}
The ground state can be obtained by the imaginary time evolution of above coupled equations.

Different from the miscible three-component droplet in  Ref.\cite{ma_borromean_2021}, in this work we focus on  a qualitatively different situation where components 1 and 3 are immiscible with strong repulsion, while (1,2) and (2,3) themselves still form binary droplets, see Fig.\ref{fig_schematic}(a). 
To achieve this in practice, we consider the boson mixture $^{23}$Na-$^{39}$K-$^{41}$K all at hyperfine state $|F=1,m_F=-1\rangle$ (denoted as 1-2-3). Near $B\sim 150{\rm G}$, we have  $(a_{11},a_{22},a_{33},a_{12},a_{13})=(52, 30, 63, -50, 213)a_0$ ($a_0$ is the Bohr radius)\cite{lysebo_feshbach_2010,viel_feshbach_2016,Schulze_2018}, and $a_{23}$ is highly tunable via a Feshbach resonance at $B_0=149.8$G with width $\Delta B\sim 25$mG\cite{tanzi_feshbach_2018}. 
One can easily check that the required mean-field instabilities, i.e., collapse for 1-2 and 2-3 and phase separation for 1-3, will occur as long as $a_{23}$ is sufficiently attractive.

Our numerical simulations  based on (\ref{GP}) indeed produce two mutually immiscible droplets, (1,2) and (2,3), as the ground state for above system, see typical density distribution in Fig.\ref{fig_schematic}(b). Remarkably, the whole system is still self-bound due to the presence of component-2. In such an immiscible phase, in principle each droplet can either stay inside as a core or outside as a shell. For the parameter regime we consider, however, the ground state is uniquely with  a higher (lower) density of 2 inside (outside), i.e., $n_2^{\rm core}>n_2^{\rm shell}$.  This can be attributed to the minimized surface energy under such configuration. For the parameter of $a_{23}$ chosen in this work, the core and shell droplets are respectively (2,3) and (1,2).

We note that similar phenomena of immiscible droplets were also found previously in Helium mixtures\cite{Pi1999,Barranco2006}, where a $^4$He droplet was coated  with a normal $^3$He liquid, and in  dipole-dipole mixtures with anisotropic density profiles\cite{bisset_quantum_2021,smith_quantum_2021}. In these studies, a long-range attraction between two droplets is required for their self-binding. In contrast, the binding of immiscible droplets in our case does not rely on any long-range force. 

\begin{widetext}

\begin{figure}[t]
    \includegraphics[width=18cm ]{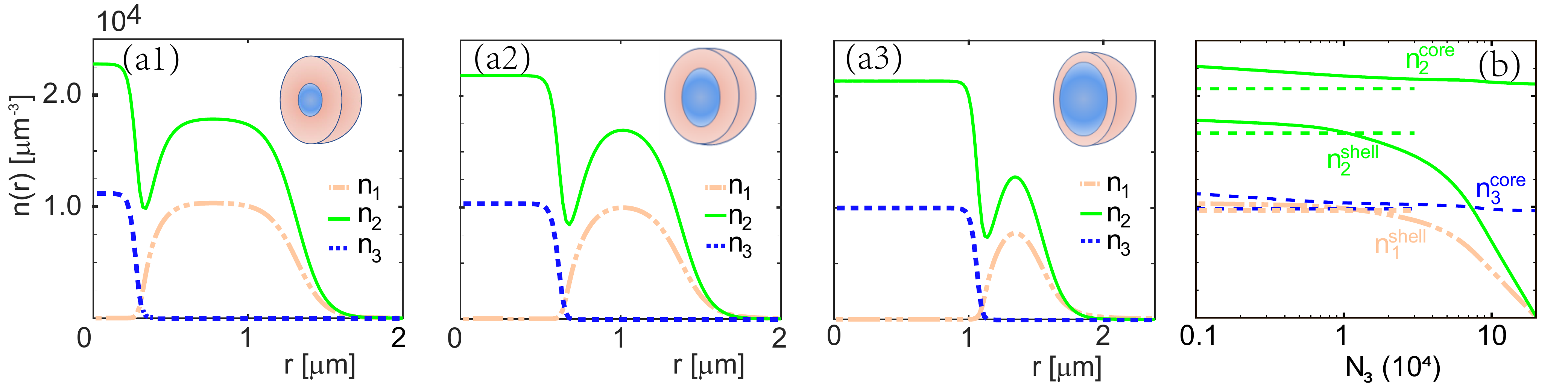}
    \caption{Expansion of shell (1,2) droplet as  increasing the size of core (2,3) droplet. Here we consider the $^{23}$Na-$^{39}$K-$^{41}$K ('1'-'2'-'3') mixture at $a_{23}=-200a_0$. (a1,a2,a3) Density profiles of ground states for different atom numbers $(N_1,N_2,N_3)/10^5=(1,1.73,0.01)$  (a1), $(1,1.86,0.1)$ (a2) and $(1,2.42,0.5)$ (a3). (b) Maximal densities of the core ($n^{\rm core}_2,n^{\rm core}_3$) and shell ($n^{\rm shell}_1,n_2^{\rm shell}$)  as functions of  $N_{\rm 3}$. 
    The dashed horizontal line with according color  denotes the equilibrium density  for the shell   ($n_i^{\rm shell}=n_i^{(0)}$)  or for the core ($n_i^{\rm core}$).}\label{fig_crossover}

\end{figure}

\end{widetext}

Now we explore the equilibrium expansion of shell droplet by increasing the core size. Let us start from the situation with a small core and a thick shell, see Fig.\ref{fig_crossover}(a1), where both droplets display flat-top densities. As increasing the size (or atom number) of the core, the shell is repelled to larger radius and becomes gradually thinner, see Fig.\ref{fig_crossover}(a2,a3); meanwhile, its flat-top profile  gradually disappears and gives way  to a Gaussian distribution with quite narrow width and low density. In Fig.\ref{fig_crossover}(b), we further extract the maximal density of each component during this process. One can see that 
the shell densities decrease rapidly as the core size increases, and eventually a very thin and dilute shell is created. 

Because of the shared component-2 as a link, the core and shell droplets, although spatially separated, are strongly correlated with each other. This is clearly  reflected in the balance condition as derived below. 
To start with, let us first consider the shell.  Since it can transfer atom or energy to vacuum, its equilibration is similar to an isolated droplet in vacuum\cite{Petrov} and determined by zero pressure and minimized energy:
\begin{equation}
 P_{\rm shell}=0; \ \ \  \epsilon_{\rm shell}=\epsilon_{\rm min}. \label{condition_0}
\end{equation}  
In the thermodynamic limit, this gives the same equilibrium density $n_i^{\rm shell}=n_i^{(0)}$ as the vacuum droplet, as well as a locked density ratio $n_i^{\rm shell}/n_j^{\rm shell}=n_i^{(0)}/n_j^{(0)}=\sqrt{g_{jj}/g_{ii}}$. However, the condition (\ref{condition_0}) does not apply to the core, since it can only transfer atoms to the shell but not directly to vacuum. In this case, the core and shell should have the same pressure and chemical potential (of the shared component-2):  
\begin{equation}
P_{\rm core}=0;\ \ \ \ \mu^{\rm core}_{2}=\mu^{\rm shell}_{2}. \label{condition}
\end{equation}
Two remarks are in order for the second condition in (\ref{condition}). First, it obviously differs from the second condition in (\ref{condition_0}), and therefore a different equilibrium density can be resulted for the core as compared to the vacuum case. Secondly, it exactly expresses the correlation between core and shell droplets. Under this condition, the core densities ($\{n^{\rm core}_i\}$) depend crucially on the shell chemical potential ($\mu^{\rm shell}_{2}$), and thus can be highly tunable by the shell parameters. 

We now analytically derive $\{n^{\rm core}_i\}$ based on (\ref{condition_0},\ref{condition}). For a thermodynamically large ($i,j$) droplet with uniform densities $\{n_i,n_j\}$, its total energy density is given by
\begin{equation}
\epsilon=\frac{1}{2}\left(g_{ii}n_i^2+g_{jj}n_j^2\right)+g_{ij}n_in_j+\epsilon_{\rm qf}. \label{E}
\end{equation}
Here $\epsilon_{\rm qf}$ is the correction from quantum fluctuations\cite{Petrov}, which determines a dimensionless function $f=\epsilon_{\rm qf} (15\pi^2/8)m_i^{-3/2}(g_{ii}n_i)^{-5/2}$. 
Then the chemical potential $\mu_j=\partial \epsilon/\partial n_j$ and pressure $P=\mu_i n_i+\mu_j n_j- \epsilon$ can be obtained straightforwardly, and the $P=0$ condition leads to the equilibrium density 
\begin{equation}
n_i=\frac{25\pi}{1024a_{ii}^3}\left(\frac{\frac{1}{2}\left(g_{jj}c_{ji}^2+g_{ii}\right)+g_{ij}c_{ji}}{g_{ii}f}\right)^2, \label{density}
\end{equation}
with density ratio $n_j/n_i=c_{ji}$.  In this way, $\mu_j$ can also be expressed in terms  of a single unknown parameter $c_{ji}$.

For the shell (1,2) droplet, the second condition in (\ref{condition_0}) results in a locked density ratio $c^{\rm shell}_{21}\approx \sqrt{g_{11}/g_{22}}$, and $n^{\rm shell}_{1,2}$ reproduce $n^{(0)}_{1,2}$ for the vacuum case\cite{Petrov}. Further, $\mu_2^{\rm shell}$ can be obtained as a function of $c^{\rm shell}_{21}$. 
For the core (2,3) droplet, by enforcing the second condition  in (\ref{condition}), and recalling $\mu_2^{\rm core}$ is just parametrized by  $c^{\rm core}_{23}$, we can then solve   $c^{\rm core}_{23}$ and further $n^{\rm core}_{2,3}$   via (\ref{density}). In Fig.\ref{fig_crossover}(b), $n^{\rm shell}_{1,2}$ and $n^{\rm core}_{2,3}$ are respectively shown by horizontal dashed and dotted lines. These analytical predictions fit well to numerical results  when the droplets are  in thermodynamical limit.

\begin{figure}[t]
    \centering
    \includegraphics[width=7.5cm, height=9.5cm]{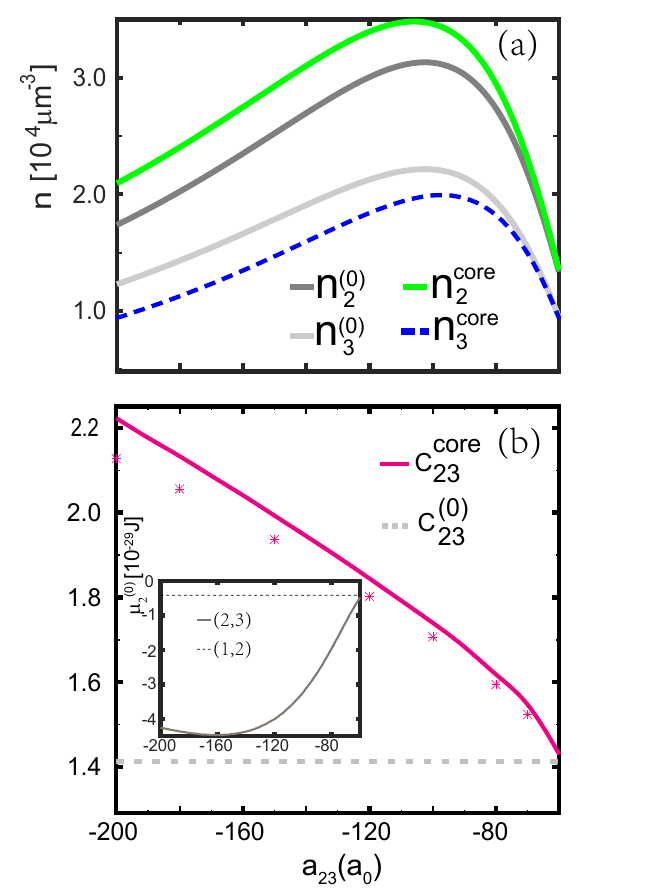}
    \caption{Tunable equilibrium densities of the core (2,3) droplet in $^{23}$Na-$^{39}$K-$^{41}$K ('1'-'2'-'3') mixture. (a) Equilibrium densities $n_{2,3}^{\rm core}$ as functions of  $a_{23}$, as compared to $n_{2,3}^{(0)}$  in vacuum. (b) Density ratio $c^{\rm core}_{23}=n_2^{\rm core}/n_3^{\rm core}$ and its vacuum counterpart $c^{(0)}_{23}=n_2^{(0)}/n_3^{(0)}$ as functions of $a_{23}$. The stars ($\star$) show numerical results by simulating the GP equations (\ref{GP}) in imaginary time for  large atom number $N_i= 10^5 \sim 10^6$. Inset  shows $\mu_2^{(0)}$  for isolated (2,3) and (1,2) droplets in vacuum.
}\label{fig_ratio}
\end{figure}

In Fig.\ref{fig_ratio} (a,b), we further show how the core densities $n_{2,3}^{\rm core}$ and their ratio $c^{\rm core}_{23}$ vary with  $a_{23}$.  One can see that both $n_{2,3}^{\rm core}$ and $c^{\rm core}_{23}$ change sensitively with $a_{23}$, and their deviations from the vacuum values ($n_{2,3}^{(0)}$ and $c^{(0)}_{23}\sim\sqrt{g_{33}/g_{22}}$) get more significant  if $a_{23}$ gets more attractive. This can be attributed to the larger mismatch of $\mu_2^{(0)}$ for isolated (2,3) and (1,2) droplets in vacuum, see the inset of Fig.\ref{fig_ratio}(b).  As a result, to balance the two chemical potentials as required by (\ref{condition}), the core has to adjust its densities to change $\mu_2^{\rm core}$, such that it can match $\mu_2^{(0)}$ in the shell. This is how the correlation is built up between core and shell, and the modified core densities are a direct evidence of such correlation. 
In this way, the core-shell structure presents a rare situation where the equilibrium density of a quantum droplet can be efficiently tuned by its surrounding environment.

\begin{figure}[t]
    \centering
    \includegraphics[height=9.5cm]{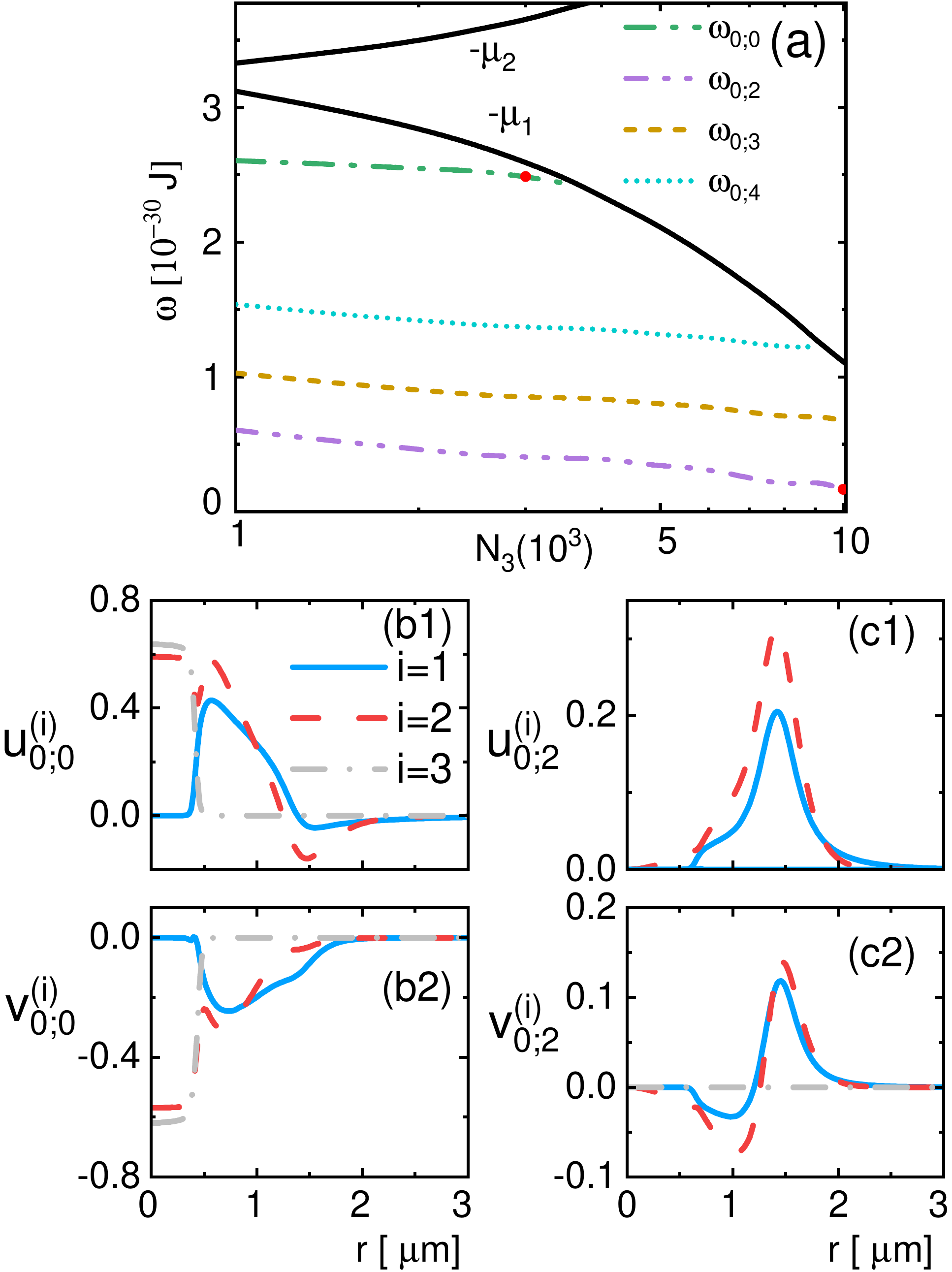}
    \caption{Collective excitations  of $^{23}$Na-$^{39}$K-$^{41}$K ('1'-'2'-'3') mixture on top of the equilibrium states in Fig.\ref{fig_crossover}(b). In (a), four lowest breathing modes $\omega_{j=0;l=0,2,3,4}$ in different angular momentum channels are shown below the atom emission threshold $\{-\mu_i\}$. (b1,b2) and (c1,c2) show the radial excitation modes $\{u^{(i)}_{0;l}(r),\ v^{(i)}_{0;l}(r)\}$ for, respectively,  $l=0$ breathing mode at  $N_3=3000$ and $l=2$ surface mode at $N_3=10^4$.}\label{fig_excitation}
\end{figure}

The core-shell correlation can also lead to unique collective excitations. Here we assume a small density fluctuation for  component-$i$:
\begin{equation}
\delta \phi_i=\exp(-i\mu_it)\sum_{j}\left(u^{(i)}_{j}\exp(-i\omega_j t)+v^{(i)*}_{j}\exp(i\omega_j t)\right).
\end{equation}
By linearizing the GP equation (\ref{GP}) in terms of $\{\delta  \phi_i, \delta  \phi^*_i\}$, we  obtain the equations for collective excitations and further solve the eigen-modes $\{u^{(i)}_{j;l}, v^{(i)}_{j;l}\}$ and eigen-energies $\{\omega_{j;l} \}$ in each  angular momentum ($l$) sector\cite{supple}.  
In Fig.\ref{fig_excitation}(a), we show the four lowest excitations modes with $j=0$ and $l=0,2,3,4$, where $l=0$ corresponds to  breathing mode and the rest three are surface modes. 
Note that the $l=1$ dipole mode is associated with the center-of-mass motion and is therefore with zero excitation energy for self-bound droplet (see also \cite{Petrov}). Here we see that as increasing the core size ($N_3$), both the breathing and surface modes gradually vanish and merge into the atom emission threshold ($-\mu_1$). This is distinct from the case without shell structure, where all excitation modes become more stable as increasing the droplet size\cite{Petrov}. To clearly see the nature of these modes, we plot out the typical radial wavefunctions $\{u^{(i)}_{0;l}(r), \ v^{(i)}_{0;l}(r)\}$ in Fig.\ref{fig_excitation}(b1,b2) and (c1,c2), respectively, for $l=0$ breathing and $l=2$ surface modes. 
Remarkably, the breathing mode displays strong core-shell correlations, where all the  three components (in both core and shell) oscillate in-phase with visible $u^{(i)}$ and $v^{(i)}$. Such correlated excitations can be attributed to the linking effect of  shared component. In comparison, the surface modes are mostly localized in the shell but occupy little in the core. In this case, the core and shell are well separated and these excitations are solely tied with the curved geometry. As the core gets larger,  the thinner shell becomes less bound and finally it cannot stabilize the surface excitations, as manifested by the disappearance of surface modes at $\omega_{0;l}=-\mu_1$.

In summary, we have demonstrated a scheme to create shell-shaped quantum droplet in ultracold boson mixtures without resorting to any trapping potential. A number of unique properties associated with the shell structure have been revealed, including its equilibrium expansion, modified core densities by core-shell balance, as well as the correlated/localized collective excitations. We have utilized $^{23}$Na-$^{39}$K-$^{41}$K mixture to demonstrate these results, while other atomic candidates may also be possible given a growing number of bosonic mixtures now available with tunable interactions\cite{K-Rb1, K-Rb, Na-Rb, K-Cs, Schulze_2018,tanzi_feshbach_2018}. In fact, quantum mixtures of three different atomic species have been successfully realized in ultracold experiments\cite{Li-K-Rb, Li-K-K}. Moreover, to facilitate the detection of shell-shaped droplet, we suggest to prepare the core-shell structure initially in an isotropic harmonic trap, with optimized mode matching to the ground state droplet\cite{guo_lee-huang-yang_2021}. In this way the system is expected to quickly relax to the shell-shaped droplet after releasing from the trap. The $l=0$ breathing mode can be naturally produced during the releasing process\cite{dynamic1, Ma2}, while the $l>0$ surface mode can be excited via certain anisotropic distortion of the cloud\cite{collective_mode_JILA, collective_mode_MIT}. The time periods of these modes crucially rely on the interaction strength, and in general a more tightly bound droplet (more attractive $a_{ij}$) will lead to a more rapid collective oscillation\cite{Petrov, supple}.

Our work opens up a new avenue to study quantum droplets with non-trivial real-space topologies, which can drive many intriguing phenomena due to the interplay with quantum fluctuations. 
One interesting subject would be the dynamical property of shell droplet, such as vortex formation and quench dynamics. Given the unique properties of self-bound droplets in vortices\cite{Li2018:TwodimensionalVortexQuantum,Zhang2019:SemidiscreteQuantumDroplets,Tengstrand2019:RotatingBinaryBoseEinstein,Caldara2022:VorticesQuantumDroplets,Gu, Oktel} and dynamics\cite{collision1,collision2,dynamic1,dynamic2,dynamic3,Ma2}, one can imagine even dramatic consequences in combination with the compactness and local curvature of shell geometry.  Moreover, it is worthwhile to study quantum fluctuation effect when the shell becomes thin enough and 
behaves as effectively 2D (where Eq.\ref{E_qf} is no longer applicable). Given that a 2D  droplet can be supported at arbitrarily low density\cite{Petrov2}, we can expect a stable shell even at very large radius with extremely thin width.  Finally, we note that the present scheme can be directly generalized to quasi-2D system in creating a toroidal droplet. More fascinating physics of these curved droplets remain to be explored in future. 

\acknowledgments
{\bf Acknowledgment.} We are grateful to Tin-Lun Ho for valuable discussions which motivated this project. The work is supported by the National Natural Science Foundation of China (12074419, 12134015, 12205365, 92476104), and the Strategic Priority Research Program of Chinese Academy of Sciences (XDB33000000).

\clearpage

\onecolumngrid
\vspace*{1cm}
\begin{center}
{\large\bfseries Supplementary Materials}
\end{center}
\setcounter{figure}{0}
\setcounter{equation}{0}
\renewcommand{\figurename}{Fig.}
\renewcommand{\thefigure}{S\arabic{figure}}
\renewcommand{\theequation}{S\arabic{equation}}

In this Supplemental Material, we provide more details on equilibrium densities and collective modes of the shell-shaped droplet.

\section*{I.\ \ \ Equilibrium densities for the core and the shell}

For a large binary ($i,j$) droplet with uniform densities $\{n_i,n_j\}$, its total energy density can be written as Eq.(6) in the main text, where the quantum fluctuation energy is expressed as:
\begin{equation}
\epsilon_{\rm qf}=\frac{8m_i^{3/2}}{15\pi^2}(g_{ii}n_i)^{5/2}f\left(\alpha,\beta,\gamma \right)
\end{equation}
with $f>0$ a dimensionless function of 
\begin{equation}
\alpha\equiv\frac{m_j}{m_i},\ \ \beta\equiv\frac{g_{ij}^2}{g_{ii}g_{jj}},\ \ \gamma\equiv\frac{g_{jj}n_j}{g_{ii}n_i}.
\end{equation}

The chemical potential $\mu_j=\partial \epsilon/\partial n_j$ can then be obtained as 
\begin{equation}
\mu_j=g_{ij}n_i+g_{jj}n_j+\frac{8\left(m_ig_{ii}n_i\right)^{3/2}}{15\pi^2}\frac{\partial f}{\partial \gamma}g_{jj},\label{mu}
\end{equation}
 and the pressure $P=\mu_i n_i+\mu_j n_j- \epsilon$ can  be obtained as 
\begin{equation}
P=\frac{1}{2}\left(g_{ii}n_i^2+g_{jj}n_j^2\right)+g_{ij}n_in_j+\frac{3}{2}\epsilon_{\rm qf}.
\end{equation}
Then $P=0$ condition leads to the equilibrium density as shown by Eq.(7) in the main text.

For the shell droplet, the chemical potential of component-2 ($\mu_2^{\rm shell}$) can be obtained via (\ref{mu}) as 
\begin{equation}
\mu_2^{\rm shell}=(g_{12}+\sqrt{g_{11}g_{22}})n_1(1-\frac{2}{3}\frac{\gamma}{f}\frac{\partial f}{\partial \gamma}).
\end{equation}
Then from the core-shell balance condition $\mu_2^{\rm core}=\mu_2^{\rm shell}$, we can solve the density ratio and further the equilibrium densities for the core.  

\section*{II.\ \ \ Collective modes}

Here we analyze the collective breathing mode of shell-shaped droplet.  Assuming a small fluctuation mode $\delta \phi_i$ for component-$i$ bosons, and only keeping the lowest-order fluctuations in the GP equation, we get 
\begin{equation}
\begin{split}
i\partial_t\delta\phi_i=&\left(-\frac{\nabla^2}{2m_i}+\sum_{j}g_{ij}n_j+\frac{\partial \epsilon_{\rm qf}}{\partial n_i}\right)\delta\phi_i+\left(g_{ii}+\frac{\partial^2 \epsilon_{\rm qf}}{\partial n_i^2}\right)(\delta\phi_i+\delta\phi_i^*)n_i\\
&+\sum_{j\neq i}\left(g_{ij}+\frac{\partial^2 \epsilon_{\rm qf}}{\partial n_i\partial n_j}\right)(\delta\phi_j+\delta\phi_j^*)\phi_i\phi_j. \label{GP2}
\end{split} 
\end{equation}
Based on the standard Bogoliubov theory, we write the fluctuation mode as
\begin{equation}
\delta \phi_i({\cp r})=\exp(-i\mu_it)\sum_{j}\left(u^{(i)}_{j}({\cp r})\exp(-i\omega_j t)+v^{(i)*}_{j}({\cp r})\exp(i\omega_j t)\right) \label{mode}
\end{equation}
where $\omega_j$ is the $j$-th eigen-mode, and $\{u^{(i)}_{j}, v^{(i)}_{j}\}$ represent according eigen-wavefunctions.  $u^{(i)}_{j}({\cp r})$ and $v^{(i)}_{j}({\cp r})$ can be further expanded in terms of the spherical harmonics as 
\begin{equation}
u^{(i)}_{j}({\cp r})=\sum_{lm} u^{(i)}_{j;lm}(r) Y_{lm}(\theta,\phi);\ \ \ \ \ v^{(i)}_{j}({\cp r})=\sum_{lm} v^{(i)}_{j;lm}(r) Y_{lm}(\theta,\phi),
\end{equation} 
with $r$ and $(\theta,\phi)$ the radial and azimuthal components of ${\cp r}$. Then (\ref{mode}) can be expanded similarly, and by plugging (\ref{mode}) into (\ref{GP2}) we find that the equations can be decoupled between different  $(l,m)$ sectors. For each given $(l,m)$, we arrive at the following equations :
\begin{equation}
  \left(\begin{matrix}
    L_1+M_1&M_{12}&M_{13}&M_1&M_{12}&M_{13}\\
     M_{12}& L_2+M_2&M_{23}&M_{12}&M_2&M_{23}\\
     M_{13}&M_{23}&L_3+M_3&M_{13}&M_{23}&M_3&\\ 
   -M_1&-M_{12}&-M_{13}&-(L_1+M_1)&-M_{12}&-M_{13}\\
   -M_{12}&-M_2&-M_{23}&-M_{12}&-(L_2+M_2)&-M_{23}\\
-M_{13}&-M_{23}&-M_3&-M_{13}&-M_{23}&-(L_3+M_3)
  \end{matrix}
  \right)
   \left(\begin{matrix}
    u^{(1)}_{j;l}(r)\\
    u^{(2)}_{j;l}(r)\\
    u^{(3)}_{j;l}(r)\\
    v^{(1)}_{j;l}(r)\\
    v^{(2)}_{j;l}(r)\\
    v^{(3)}_{j;l}(r)
  \end{matrix}
  \right)
  =\omega_{j;l}\left(\begin{matrix}
    u^{(1)}_{j;l}(r)\\
    u^{(2)}_{j;l}(r)\\
    u^{(3)}_{j;l}(r)\\
    v^{(1)}_{j;l}(r)\\
    v^{(2)}_{j;l}(r)\\
    v^{(3)}_{j;l}(r)
  \end{matrix}
  \right),   \label{breathing}
  \end{equation}
where
\begin{equation}
\begin{split}
L_i&=-\frac{1}{2m_i}\left(\frac{1}{r} \frac{\partial^2}{\partial r^2}r-\frac{l(l+1)}{r^2}\right)+\sum_jg_{ij}|\phi_j|^2+\frac{\partial \epsilon_{\rm qf}}{\partial n_i}-\mu_i\\
M_i&=\left(g_{ii}+\frac{\partial^2 \epsilon_{\rm qf}}{\partial n_i^2}\right)n_i\\
M_{ij}&=\left(g_{ij}+\frac{\partial^2 \epsilon_{\rm qf}}{\partial n_i\partial n_j}\right)\phi_i\phi_j.\\
\end{split}
\end{equation}
and the normalization condition is
\begin{equation}
\int d{\cp r}\sum_{i=1}^{3}\left(|u^{(i)}_{j;l}|^2-|v^{(i)}_{j;l}|^2\right)=1. \label{nor}
\end{equation}
Here since all the $m$-solutions are degenerate, we have dropped this index for simplicity.  By solving above equations, we can obtain the eigen-energy $\omega_{j;l}$ and eigen-wavefunctions  $\{u^{(i)}_{j;l}(r), v^{(i)}_{j;l}(r)\}$ for collective excitations in each $l$ channel. The breathing modes correspond to the solutions with $l=0$, and all higher $l(>1)$ ones are surface modes. Note that for the self-bound droplet, the center-of-mass motion causes no excitation energy and thus the lowest dipole mode (with $l=1$) has zero eigen-energy.

\begin{figure}[h]
    \centering
    \includegraphics[width=10cm]{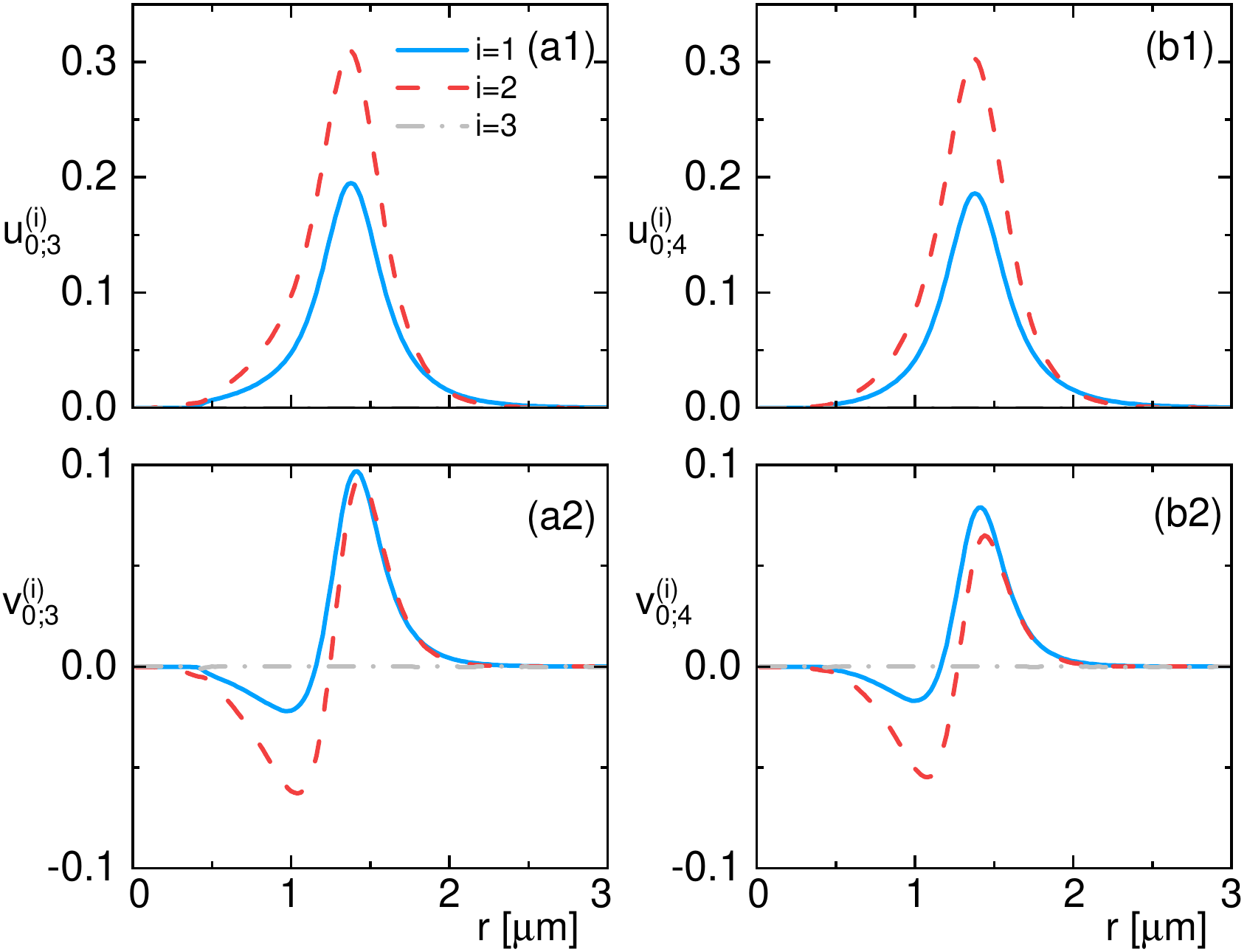}
    \caption{(a1,a2) and (b1,b2) show the radial excitation modes $\{u^{(i)}_{0;l}(r),\ v^{(i)}_{0;l}(r)\}$ for, respectively,  $l=3$  and $l=4$ surface modes at $N_3=3000$. }\label{fig_s1}
\end{figure}

In Fig.4 of the main text, we have shown the typical $\{u^{(i)}_{0;l}(r), v^{(i)}_{0;l}(r)\}$ for $l=0$ breathing mode and $l=2$ surface mode. In Fig.\ref{fig_s1} we further plot out typical $\{u^{(i)}_{0;l}(r), v^{(i)}_{0;l}(r)\}$ for surface modes with higher $l=3,4$. One can see that all these modes are  localized within the shell while with little occupation in the  core. This means that the surface excitation in the core is strongly suppressed by the shell structure. 
 
We have also checked that for very large core and very thin shell, these surface modes are no longer stable  and the whole system keeps emitting particles to the vacuum under the surface excitations. In this regime, all $\omega_{0;l}$ stay above the atom emission threshold $-\mu_1$. Physically, it is because the gradually thinner shell is less bound (with increasing $\mu_1$  or decreasing $-\mu_1$), and therefore it cannot afford any surface excitation (with $\omega_{0;l}+\mu_1>0$). This directly shows the crucial role of shell structure in modifying the collective exciations in quantum droplets.

For all the plots of collective modes in this work, we have considered the $^{23}$Na-$^{39}$K-$^{41}$K ('1'-'2'-'3') mixture at (tunable) $a_{23}=-200a_0$. The resulting breathing mode $\omega_{0;0}$, as shown in Fig.4(a) of the main text, is $\sim 2.5\times 10^{-30}J$. This corresponds to a quite short oscillation period $T=h/\omega_{0;0}\sim 0.25ms$. To increase the period, one can tune $a_{23}$ to be less attractive. 
This can be qualitatively understood from the scaling analysis of binary droplet in vacuum, where all the energies are scaled by $\xi^{-2}\sim |\delta g|^{3}$, with $\delta g$ the effective mean-field attraction. Therefore, a more attractive $a_{23}$ (giving a larger $|\delta g|$) will lead to a larger energy scale and thus  more rapid collective oscillations. We have confirmed this in our numerics: if change  $a_{23}$ to $-60a_0$, we will have $\omega_{0;0}\sim 0.6\times 10^{-30}J$ at typical numbers $N_1=N_3=10^5$ (with adjusted optimal $N_2=3.13*10^5$), giving a longer period $T\sim 1ms$. (Note that in this case we need a larger $N_3$, as compared to the case of  Fig.4(a), in order to support the $(2,3)$ droplet with less attractive $a_{23}$.)

Different from $l=0$ breathing mode, $l>0$ surface modes are only localized within the shell and therefore insensitive to the change of $a_{23}$. Moreover, for the parameters taken in Fig.4, these modes have much smaller oscillation frequencies and thus longer periods than the breathing mode. For instance, for $l=2$ surface mode shown in Fig.4(c1,c2), we have  $\omega_{2,0}\sim 0.15\times 10^{-30}J$, corresponding to a period $T=h/\omega_{2;0}\sim 4.5ms$.



\end{document}